# Scaling of Laser Produced Pair Plasmas And Comparison With Experimental Data

Alexander Henderson, Edison Liang, Pablo Yepes (Rice University, Houston, TX, 77005), Hui Chen, Scott Wilks (Lawrence Livermore National Laboratory, Livermore, CA, 94550).

Abstract: We report simulation results of pair production by ultra-intense lasers irradiating a gold target using the GEANT4 Monte-Carlo code. Certain experimental features of the positron and electron energy spectra are reproduced, as well as trends with regard to target thickness and hot electron temperature $T_e$. For $T_e$ in the range 5-10 MeV, the optimal target thickness for pair production is found to be about 3 mm. Further Monte-Carlo simulations may aid in the optimization of laser-driven positron sources.

I: Introduction

When a laser of intensity greater than about $10^{18}$ W/cm$^2$ strikes a solid target it couples a significant fraction of its energy to hot electrons with energy $> m_e c^2$ [1,2,3,4]. If the target is a high-Z material such as gold these hot electrons will create electron-positron pairs via two possible channels: the (electron) trident process[5] and the Bethe-Heitler process[5]. The trident process produces electron-positrons pairs directly via interaction with the nucleus (eZ -> eZe$^+$e$^-$)[5]. The Bethe-Heitler process produces pairs by first producing photons via bremsstrahlung radiation (eZ -> eZg), then the photon interacts with a nucleus to produce a pair (gZ -> Ze$^+$e$^-$)[5]. The outgoing jet of electrons and pairs will then constitute a non-neutral pair plasma. Positron creation through these channels via laser-matter interactions was first demonstrated experimentally by Cowan et al.[6], and was further explored by Gahn et al[7]. But the major breakthrough came when Chen et al created copious pairs using the Lawrence Livermore National Laboratory (LLNL) Titan laser.[8,9] The positron density of such a pair plasma can potentially exceed $10^{18}$ per cubic centimeter[1], higher than any pair plasmas produced by other laboratory-based methods. High density pairs have many potential applications, from laboratory astrophysics to antimatter creation[10].

Here we report Monte Carlo simulation results of pair creation using the GEANT4 version 4.9.4[11,12], and compare our results with data taken on the Titan laser at Lawrence Livermore National Laboratory (LLNL)[8,9]. These simulations idealize the incident hot electron distribution by a Maxwellian of a given temperature instead of using output spectra of particle-in-cell (PIC) simulations. Despite such approximations, we obtain high energy electron spectra that are in good agreement with data.

II: Experimental Methods and Setup

GEANT4 is a Monte-Carlo high energy physics code designed to track particles with energies from the multiple GeV range down to the sub-keV range. Its basic process is to track particles through matter and follow their evolution through all electromagnetic and hadronic interactions they may undergo. A variety of physical processes may be implemented such that at each step the tracked particle's probability of interacting via that process is either calculated via a formula or taken from a table. Using random numbers GEANT4 then determines where that interaction takes place and changes the particle's state accordingly. Daughter particles are tracked in the same manner, allowing simulation of cascades. Figure 1 illustrates an electron cascade simulated with GEANT4 using Bethe-Heitler, bremsstrahlung, and Compton scattering processes. The formulas and tables used for calculating electromagnetic cross-sections have been shown to provide an accuracy in the lower energy ranges, on the order of 10 MeV and lower, better than the Electron Gamma Shower (EGS) code[13].

In an experiment conducted on Titan[8,9], emergent electrons and positrons were collected at an angle of 28 degrees to the target normal. See figure 2 for a rough diagram of the experimental set-up. Note that in order to reduce processing time in the simulation , the region comprised of concentric rings to either side of 28 degrees with respect to the target normal was used for the detection region rather than a box. In the Titan experiment, the laser was polarized into the target-detector plane (s-polarized), so for the purpose of simulating the detector in that experiment, this assumption of axial symmetry is justified. For simplicity, the simulation fired electrons in a cone with axis parallel to the target normal. Chen et al.[14] indicate that the angle of laser incidence does have some effect. Specifically, a laser with an incidence angle of 18 degrees with respect to target normal resulted in the angular distribution of positrons being centered near -10 degrees from the target normal on the OMEGA laser. However, the Titan laser had smaller focal spot size with respect to the target size[14] which should reduce the effect enough that the lower angular resolution in simulation (+ or – 2.5 degrees about the desired measure) should overwhelm the angle offset.

For our simulation, we incorporated the Compton, Photoelectric, gamma-to-e+/e- pair creation, Rayleigh scattering, bremsstrahlung, and pair annihilation processes provided in the GEANT4 code[11,12,15]. For the electron energies in the MeV range, the Trident process will be negligible compared to the Bethe-Heitler process for targets thicker than a few hundred microns[16]. Consequently, the Trident process is ignored for simulations in this range. In addition, GEANT4 does not simulate plasma effects, but those effects will be addressed separately later. The temperature and angular spread of the laser-produced incident hot electrons were taken as free parameters and obtained by fitting simulated output to electron data from the Titan experiments. The simulation incident electrons were fired flush with the edge of a gold target, and emergent positrons, electrons, and gamma rays were detected on a large slab positioned far from the target. The incident electrons were spread over a disc 8 micrometers in diameter and with fixed Gaussian angular and time distributions (see section IV). We vary the target thickness and incident hot electron temperature kT in the current studies.

III: Scaling Studies with Thickness and Temperature

In the GEANT4 simulations we observe how pair outputs scale with thickness at given hot electron temperatures. We first look at the emergent positron-to-electron ratio as a function of thickness. We can do this in two ways: first, we can study the ratio of emergent positrons to emergent electrons, which is a measure of the pair "purity" of the outflow plasma. This can be compared with experiment once we allow for positron spectrum shift due to the effects of the electric field that the ions and electrons form (the sheath field) and consider only the energy range for which the simulation accurately reproduces the shape of the spectra (see section IV). For some experimental runs, positron data was available but electron data was not. The available electron data was extrapolated to provide an estimate of the number of electrons at those temperatures. For targets which are not too thick, the emergent electrons should be dominated by the initial electrons which remain after passing through the target. Hence, the number of emergent electrons should roughly decay exponentially with thickness, since the initial number of electrons should be similar. Due to the inherent variation in laser parameters from shot to shot, this will not be exact, but it will at the very least provide an order-of-magnitude estimate, which is sufficient for our purposes. Using the available data, the best-fit exponential distribution was $N = 3.13*10^{10}*\exp(-1.1*T)$ where N is the number of electrons remaining and T is the target thickness in millimeters (mm).

The second thing we can study is the ratio of the energy in the emergent positrons to incident energy (the "yield"). The energy incident on the target couples to the energy of the hot electrons, which is the quantity relevant to the simulation. Thus in effect the simulation fixes these values. Hence, to find the "yield" from simulation in a manner than can be compared with Titan data, we scale the ratio of emergent positrons to incident electrons by the ratio of the incident electron temperature to some baseline temperature. Then, we scale the number of emergent positrons in the Titan data to some arbitrary number chosen so that the "yield" thus obtained is close to simulation at some reference thickness, such as 1 mm. This creates a valid comparison because the positron spectra match almost exactly between data and simulation, once a sheath field has been added to the latter (see Section IV). Thus, after all of this we obtain an estimate of the relative scaling of positron "yield" with thickness for the hot electron temperatures, and thus the laser energies, given, rather than making an absolute comparison.

We first consider the outflow e+/e- ratio. We do not take the ratio over the whole detected energy range. Due to the failure of the simulation to reproduce accurately spectral features in the range below 4 MeV (see Section IV), we only take the energy range between 4 MeV and 40 MeV. The latter value is the lowest energy where the signal is distinguishable from noise in all available Titan data. Note that previous published Titan results[8,9] used a lower energy cut-off of 0.8 MeV, encompassing most of the region of concern. Consequently, the values from Titan data here differ somewhat from those published in Chen et al. 2009[8,9]. The simulation shows a quadratic increase in the e+/e- ratio with thickness up to a certain point, dependent on the input temperature, after which it levels out and approaches 1 asymptotically (see figure 3a). This is expected since for very thick targets the incident electrons are mostly depleted and only newly created pairs can emerge. Two conclusions may be drawn from this: at higher thicknesses the ratio is relatively insensitive to initial electron temperature, but at lower thicknesses it is highly dependent on the electron temperature. Comparisons with observed data require applying a sheath electric field to shift the positron spectrum, which has been done here and will be discussed more in Section IV. In the simulation, the e+/e- ratio increases roughly exponentially with thickness up to 5 mm, as expected. Presumably if this process were carried out to higher thicknesses it would level out as the number of hot electrons that emerge approaches 0 and produced e+/e- pairs become significant, but the signal-to-noise ratio is too low at thicknesses above 5 mm for reasonable comparisons. In Titan[8,9] experiments, the increase appears to still be exponential, but to have a different rise constant. At 5 mm thickness, the electron data was not available, so the electron number is based on the exponential extrapolation mentioned above. The ratio obtained in this manner is greater than 1, which is unreasonable, so we conclude that already at 5mm the pair-produced electrons make a significant contribution in terms of number of electrons present. Were this adjustment applied, the ratio obtained from the data would come down. Since the Monte-Carlo simulation itself is intended to obtain the data required to make this adjustment, we cannot make further comparison between the data and simulation at 5 mm. Discrepancies at other thicknesses can be attributed to the inexact method of simulation of the sheath field and so will be discussed further in Section IV.

As noted above, in order to make comparisons with the experimental data we must shift the simulated positron spectrum up to match the observed spectra. We do this by applying a simple electric field in the space between the target and the detector. Note that this simple method is only expected to work over the whole observed energy range for the positrons, which do not have an observable number of particles at higher energies. The emergent electrons, as noted earlier, are themselves the sheath field generators and furthermore do have a significant number of particles at higher energy. Hence, some of them pass to the detector before the field has been set up. However, these particles are mostly in the tail of the spectrum, so with the upper energy cut-off of 40 MeV used here, the effect on the e+/e- ratio should be small. As a result, the field in question is set up very rapidly with respect to the passage of the positrons[14]. Keeping the above caveat in mind the field does not need to be time-dependent. The sheath potential is determined empirically by the relative positions of the positron peaks (see Section IV). This applies reasonably well for all thicknesses observed. Note that the field did not follow a predictable progression since the field strength is dependent not only on target thickness but also on laser intensity and laser energy, which have some variability. Thus the field had to be calculated on an ad hoc basis, and hence simulation results which are not directly compared with data, such as figure 3a do not include the sheath field.

We next discuss the yield results. According to figure 4a, the e+ yield per incident hot electron increases with thickness to an absolute maximum value which is dependent on temperature. Above this point, it slowly falls off. At higher temperatures, there is also a second local maximum in the yield which is slightly below the absolute maximum yield. In either case, the maximum yield increases with effective incident electron temperature. We note that the improvement in maximum yield suffers diminishing returns as the temperature is increased: increasing the effective temperature from 5 MeV to 10 MeV increases the e+/incident e- ratio from $3.4 \times 10^{-3}$ to $3 \times 10^{-2}$, but increasing from 10 to 15 MeV only increases it further to $3.4 \times 10^{-2}$. We note in passing that the simulation predicts that the gamma yield matches the positron yield very closely, except for a scaling factor. We can compare positron yield directly with experimental data by applying the sheath field as mentioned above. No such experimental data was available for gamma yield. If we look at figure 4b, we can see that there is a qualitative similarity in the shape of the trends in yield. Specifically, the relative changes in yield between 1 mm and 2 mm targets and between 3 mm and 5 mm targets are similar. However, it is also clear that calibrating the yield at one pair of thicknesses would cause the predictions for the other pair to be inaccurate. Possible reasons for this will be discussed in Section VI.

IV: Spectra Comparison with Titan Data:

We vary the input electron temperatures for each target to obtain a "best fit" match with the measured electron spectrum from Titan experimental data. In particular, we match the slope of the high-energy electron tail in a log-linear plot. Pondermotive scaling[17] gives an expected range of 1-4 MeV for incident hot electron temperature, but it is already known[7] that this is an underestimate, so temperatures were varied starting from 5 MeV and adjusted until a match was obtained. In this simulation, temperature differences smaller than 1 MeV resulted in differences in output that were only barely discernible, and hence adjustments were made in units of 1 MeV.

For the available Titan data, the simulations give a best-fit incident electron temperature ranging from 5 to 7 MeV, in general agreement with Chen et al.[8,9]. For the purpose of comparing with experimental data we use the spectra gathered near 28 degrees from the target normal, the position of the detector in the Titan experiment.

From comparisons in figure 5a, we see that the electron spectrum at 1mm fits the high-energy regime best with an incident electron temperature of 6 MeV. With this temperature, the spectrum compares well with the spectrum from Titan data for energies above 4 MeV. For lower energies, we see that the simulation drops down, experiencing a narrow spike at very low energies due to Compton electrons[18]. We note that earlier EGS simulations showed no such spike[8,9] (see figure 5a). The Titan data shows a much sharper uptick at low energy than the simulation, ending at about 4 MeV. While part of this spike is due to these Compton electrons the rest of the low energy behavior cannot be reproduced with GEANT4. Hence, we anticipate that this lower energy increase in the experimental data is due to the effects of the charged plasma outflow and sheath electric field. This means that one cannot simply adjust the electron spectrum by means of a constant linear electric field, but must take the plasma dynamics into consideration. This will naturally shift some electrons to lower energy, but this will occur in a more complex manner than the shift of the positrons to higher energy. We speculate that a PIC simulation of the sheath field process would resolve this discrepancy, but this falls outside the scope of this paper. Besides this spike, there are other discrepancies. This is discussed further below. Electron spectra at 2mm and 3mm are similar: the experimental spectra have a low-energy spike which is only partly reproduced by GEANT4 simulation. Full experimental electron spectra were unavailable at higher thicknesses, where the signal-to-noise ratio is much lower and the spectra correspondingly less clear.

The simulated positron spectra match the Titan data extremely well in shape at all examined thicknesses. There are slight discrepancies at low energies, where there is a slight upturn in the experimental data which reduces with thickness, becoming a slightly less steep part of the spectrum at 3mm. For simplicity, only the 1 mm and 3 mm comparisons are shown, the other thicknesses matching the general pattern of one of these two (figure 6). Note that the entire positron spectrum can be shifted by the application of a simple electric potential to obtain a good match for most of the spectrum. Since the electrons and not the positrons are the field generators, and since the field is set up quickly with respect to the time it takes for positrons to reach the detector, this is a reasonable conclusion. The location of the positron peak, and thus the strength of the electric field required to match the positron spectrum, varies based on the cross-section of the target[14]. After this application, the simulated positron spectra all match experiment very closely in both shape and position. Finally, we note that at the lowest thickness studied in this experiment, 0.38 mm, there are not enough pairs produced in a target of this thickness to get a good positron signal above 6 MeV. Consequently, this thickness was excluded from consideration.

Due to the presence of the low energy "spike" in the electron spectra data (figure 5), we compare the data and simulation for energies excluding this feature (see Section II). Specifically, since the spike ends at 4 MeV for a 1 mm thick target, and at or below 4 MeV for other thicknesses, comparison is made in the range from 4 to 40 MeV, as stated before.

Taking all these things into account, we can see that the electron spectra match fairly well in this energy range for an electron input temperature kT of 6 MeV and a target thickness of 1 mm (see figure 5a). The data possesses a flatter peak than the simulation, but this could be due to plasma sheath and other effects not included in the Monte Carlo simulations.

V: Angular Distributions:

Next we consider the angular distribution of the positrons and electrons. We used a Gaussian injection cone with an opening angle with a Full Width Half Maximum (FWHM) of 20 degrees, which results in emergent e+ and e- angular distributions with FWHMs of around 50 degrees with no sheath field applied at a thickness of 1 mm (figure 7). This indicates a widening of the output cones compared with the e- input cone, which is expected.

At a given thickness, the electron angular distribution narrows with increasing temperature as expected. For simplicity, we examine them without the sheath field adjustments since we already have some idea of what those do. For instance, at 1 mm, it goes from a FWHM of around 56 degrees at a temperature of 5 MeV to a FWHM of around 48 degrees at a temperature of 7 MeV (figure 7a). The positron spectrum also narrows, going from a FWHM of about 64 degrees at a temperature of 5 MeV to a FWHM of about 40 degrees at a temperature of 7 MeV (figure 7b). These results are expected since more higher-energy electrons are present with higher initial temperatures, both in that a higher fraction of the initial electrons are at higher energies and the electrons produced will have higher energies on average. In addition, many of these produced electrons would not have been present at lower initial temperatures. The incident electrons will tend to have smaller angles-of-flight on average, and hence so will pairs produced from them. The effect is more severe in the positrons since they are all secondary targets.

VI Discussion:

From above, we see that the simulation, after appropriate adjustments, successfully reproduced the positron spectra. It also successfully reproduced some qualitative trends in the positron yield. However, this GEANT4 simulation did not correctly reproduce some observed features of the positron yield, nor did it correctly reproduce the positron-to-electron ratio at certain target thicknesses. These discrepancies remain even after some simplistic adjustments are made to account for the sheath field. The discrepancy in ratio occurs in a similar manner for both GEANT4 and EGS simulations.

Tests in Amako et al.[19] on GEANT4 indicate that the error in the packages used should not exceed 5% in the relevant energy range, not nearly enough to account for the observed discrepancies with Titan data, so the cascade processes themselves may be excluded as a source of error. The plasma and geometric effects inherent in a real sheath field likely account for some of the error in ratio and yield.

These effects would alter the angular distribution somewhat if properly simulated, and would affect the positron and electron angular distributions differently. Hence, the differences in ratio for 1, 2, and 3 mm targets may well be explained by these effects. At 5 mm, we have additionally that the assumptions used to estimate the electron count break down, as mentioned in Section III. Since the sheath potential is much lower at 5 mm, sheath field effects are expected to be less significant to the discrepancies at that thickness. More detailed sheath field effects may also resolve the discrepancies in the electron spectra below 4 MeV. It is also possible that the assumption of a single Maxwell distribution for the incident electron energies is inaccurate enough at energies below 4 MeV to significantly impact the electron spectra in that region. There is some possibility that energy conservation considerations and the electron counter-current required by them have some effect, but with a high-Z material such as gold the effect on streaming particles should be negligible.[20] These discrepancies and shortcomings will be explored in future work.

In summary, GEANT4 Monte-Carlo simulation of pair production by hot electrons passing through a gold target shows general agreement with Titan data[8, 9], although some significant discrepancies remain. Simulation shows positron yield peaking with thickness between 3 and 4 mm. This is not precisely consistent with the data, but the data is sparse enough that it is difficult to tell if this discrepancy is significant. Simulation also predicts that the positron yield increases with temperature at least up to 15 MeV. In addition, simulation predicts the e+/e- ratio in the outflow increases with both hot electron temperature and with target thickness. As the thickness grows, the e+/e- ratio coming off of the target is predicted to approach 1, though in actual practice this would be difficult to measure as the emergent positron and electron signals decrease towards the noise threshold as thickness increases. Future work will focus on using this simulation to design experiments and using future available data to test its usefulness.

Acknowledgments.

This work was supported by DOE-SC0001481 and NSF 0909167 at Rice University. At LLNL this work was performed under the auspices of USDOE under contract number DE AC52-07NA27344.

Bibliography

[1] E. Liang, S. Wilks, M. Tabak, Physical Review Letters 81, 4887 (1998).

[2] S. C. Wilks, W. L. Kruer, M. Tabak, and A. B. Langdon. Physical Review Letters. 69 (9), 1383-1386 (1992).

[3] Donald Umstadter. Phys. Plasmas 8, 1774 (2001).

[4] P Gibbon and E Förster. Plasma Phys. Control. Fusion 38, 769 (1996).

[5] W. Heitler, *The Quantum Theory of Radiation, 3rd Edition* (Oxford University Press, London 1954) p. 256-265.

[6] T. E. Cowan et al., Laser Part. Beams 17, 773 (1999).

[7] C. Gahn et al., Phys. Plasmas 9, 987 (2002).

[8] Hui Chen, Scott C. Wilks, James D. Bonlie, Edison P. Liang, Jason Myatt, Dwight F. Price, David D. Meyerhofer, and Peter Beiersdorfer. Physical Review Letters. 102, 105001 (2009).

[9] Hui Chen, S. C. Wilks, J. D. Bonlie, S. N. Chen, K. V. Cone, L. N. Elberson, G. Gregori, D. D. Meyerhofer, J. Myatt, D. F. Price, M. B. Schneider, R. Shepherd, D. C. Stafford, R. Tommasini, R. Van Maren, and P. Beiersdorfer. Phys. Plasmas 16, 122702 (2009).

[10] E. Liang, High Energy Density Physics, 6. p. 219-222 (2010).

[11] Agostinelli S *et al* 2003 GEANT4–a simulation toolkit *Nucl. Instrum. Meth. A* **506** 250–303

[12] Allison J *et al* 2006 GEANT4 developments and applications *IEEE Trans. Nucl. Sci.* **53** 270-8.

[13] M. Vilches, S. Garcia-Pareja, R. Guerrero, M. Anguiano, and A. M. Lallena. Nuclear Instruments and Methods in Physics Research B. 254, 219-230 (2007).

[14] Hui Chen, S. C. Wilks, D. D. Meyerhofer, J. Bonlie, C. D. Chen, S. N. Chen, C. Courtois, L. Elberson, G. Gregori, W. Kruer, O. Landoas, J. Mithen, J. Myatt, C. D. Murphy, P. Nilson, D. Price, M. Schneider, R. Shepherd, C. Stoeckl, M. Tabak, R. Tommasini, and P. Beiersdorfer. Physical Review Letters. 105, 105003 (2010).

[15] http://geant4.web.cern.ch/geant4/UserDocumentation/UsersGuides/PhysicsReferenceManual/fo/PhysicsReferenceManual.pdf retrieved 11 January 2011.

[16] Ken'ichi Nakashima and Hideaki Takabe. Physics of Plasmas. 9(5), 1505-1512 (May 2002).

[17] J. Myatt, J. A. Delettrez, A. V. Maximov, D. D. Meyerhofer, R. W. Short, C. Stoeckl, and M. Storm. Physical Review E. 79, 066409 (2009).

[18] George B. Rybicki and Alan P. Lightman, *Radiative Processes in Astrophysics* (John Wiley and Sons, Inch, New York 1979).

[19] Katsuya Amako, Susanna Guatelli, Vladimir N. Ivanchenko, Michel Maire, Barbara Mascialino, Koichi Murakami, Petteri Nieminen, Luciano Pandola, Sandra Parlati, Maria Grazia Pia, Michela Piergentili, Takashi Sasaki, and Laszlo Urban. IEEE Transactions on Nuclear Science. 52(4),

910-918 (August 2005) .

[20]A R Bell, J R Davies, S Guerin, and H Ruhl. Plasma Phys. Control. Fusion 39, 653 (1997).

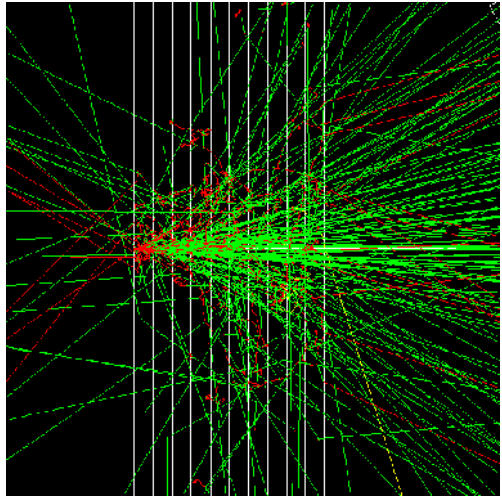

Figure 1: Diagram of a set of simulation shots. Electrons are red, positrons are yellow, and gamma rays are green. Note positron leaving target at steep angle.

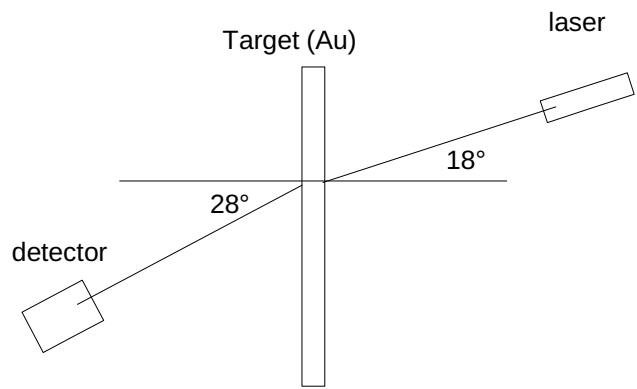

Figure 2a: Titan experimental set-up.[11,12]

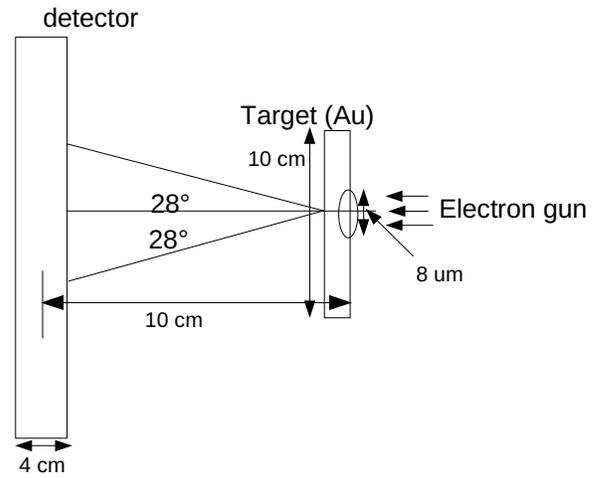

Figure 2b: GEANT4 Simulation set-up. Electrons are shot flush with the edge of the target. Data for a given angle is taken along a circle on the large slab detector.

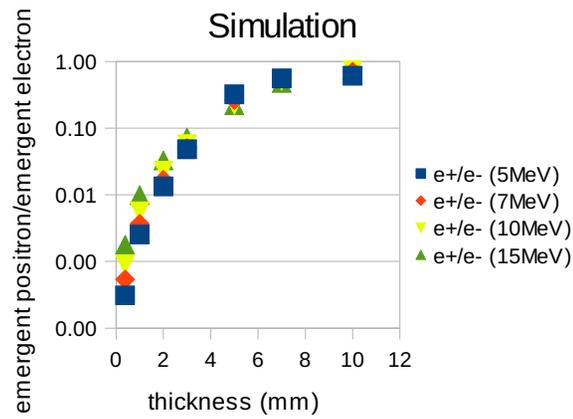 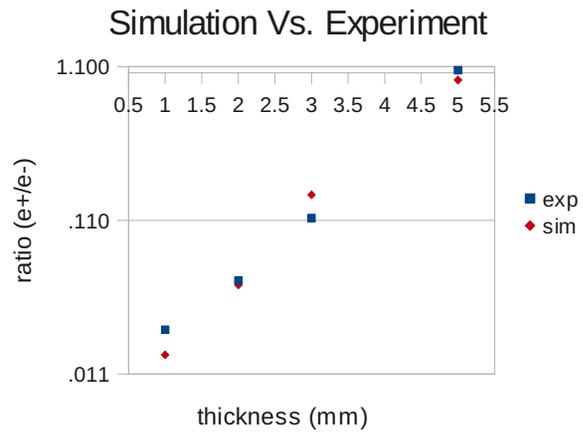

Figure 3a: Simulated e+/e- ratio in outflow, without sheath field.

Figure 3b: Comparison between simulation with sheath field and Titan experimental e+/e- ratio. Hot electron temperatures are the best fits from simulation; values are noted in text.

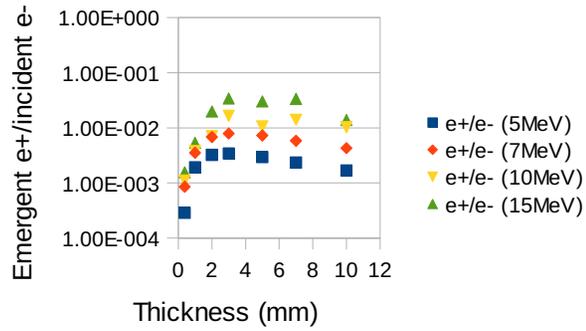

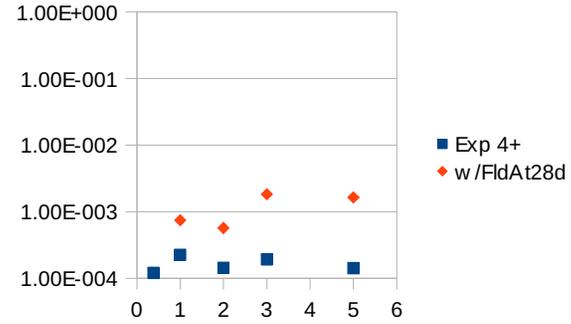

Figure 4a: Simulated yield according to simulation versus thickness at various energies.

Figure 4b: Yield vs. thickness comparison between experiment (blue) and simulation (red). The vertical scale is arbitrary. The hot electron temperature is lower for 2 mm than other thicknesses (see text).

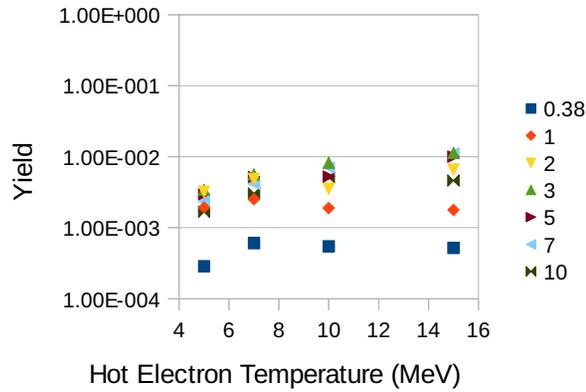

Figure 4c: Simulated yield versus incident electron Temperature at various thicknesses in mm. This is a reorganization of 4a.

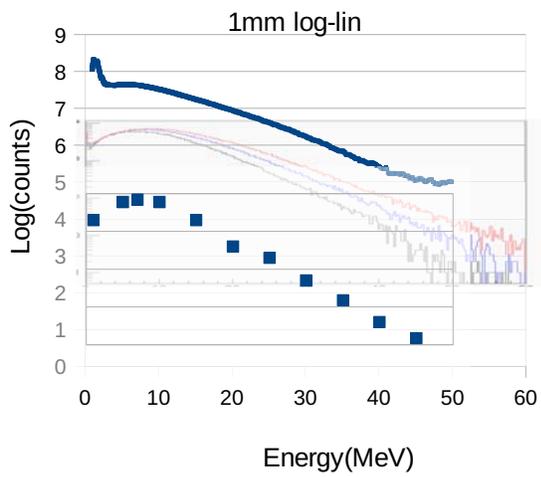 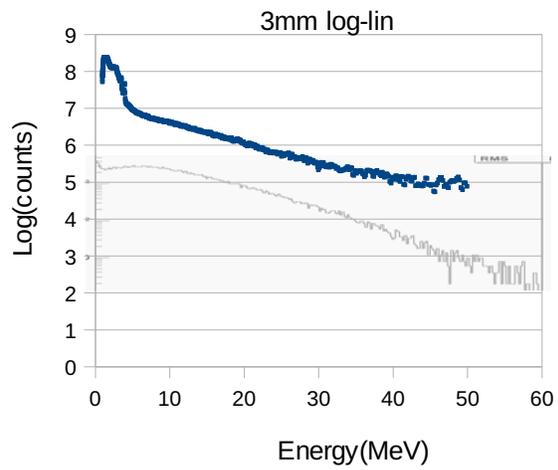

Figure 5a: Comparison between electron spectrum from data (top) and simulation with GEANT4 at 5 MeV (black thin line), 6 MeV (blue), and 7 MeV (red). The bottom set of points is an EGS simulation based on a graph from Hui Chen et. al.[11, 12] The simulation spectra are rescaled for easy comparison.

Figure 5b: Comparison between electron spectrum from data (top) and simulation with GEANT4 at 7 MeV (black thin line). The simulation spectrum is rescaled for easy comparison.

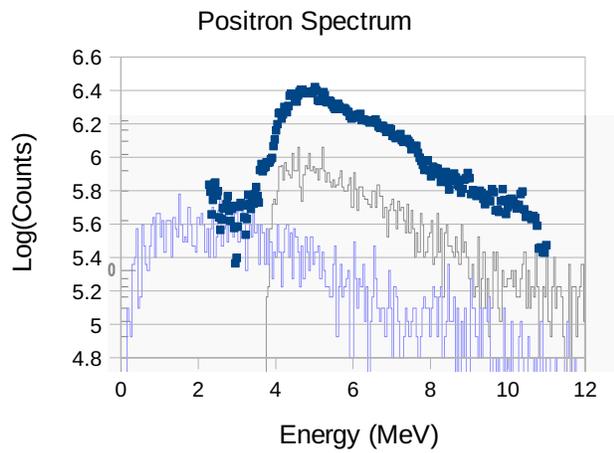
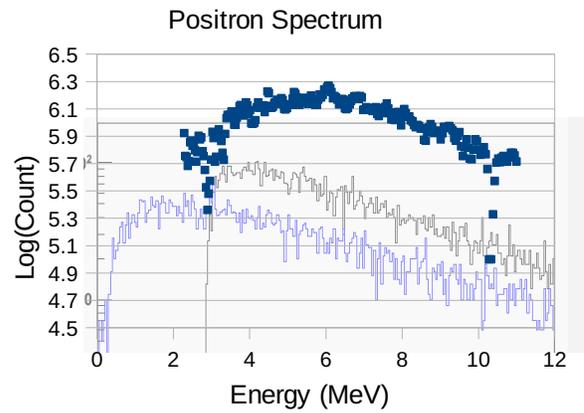

Figure 6a: Positron spectrum comparison at 1 mm. Top is experimental data, blue is GEANT4 simulation without a sheath field, and black is GEANT4 simulation with a sheath field. Simulation data is rescaled for comparison purposes.

Figure 6b: Positron spectrum comparison at 3 mm. Top is experimental data, blue is GEANT4 simulation without a sheath field, and Black is GEANT4 simulation with a sheath field. Simulation data is rescaled for comparison purposes.

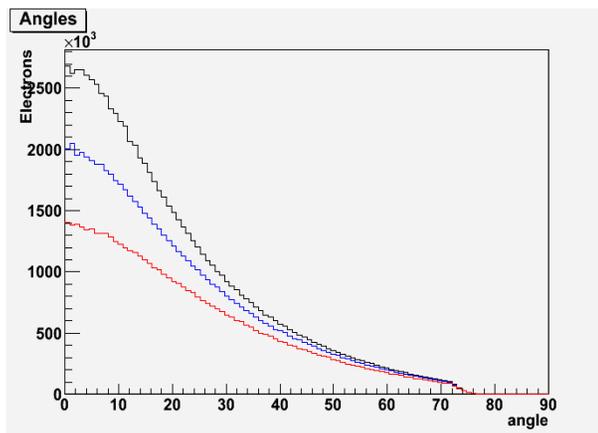 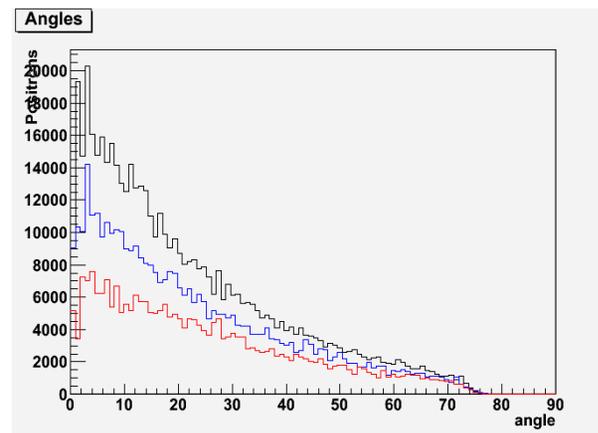

Figure 7a: Simulated emergent electron angular distribution (electrons per steradian) for a 1 mm target with incident electron temperatures of 5 MeV (red), 6 MeV (blue), and 7 MeV (black). The detector geometry causes an artificial drop-off at around 72 degrees.

Figure 7b: Simulated emergent positron angular distribution (electrons per steradian) for a 1 mm target with incident electron temperatures of 5 MeV (red), 6 MeV (blue), and 7 MeV (black). The detector geometry causes an artificial drop-off at around 72 degrees.